\documentclass[onecolumn,amsmath,amssymb]{revtex4}

\usepackage{graphicx,graphics,multirow,caption,tabularx}
\def\unit{\leavevmode\hbox{\small1\kern-3.6pt\normalsize1}}
 \normalsize

\def\lsim{\raise0.3ex\hbox{$\;<$\kern-0.75em\raise-1.1ex\hbox{$\sim\;$}}}
\def\gsim{\raise0.3ex\hbox{$\;>$\kern-0.75em\raise-1.1ex\hbox{$\sim\;$}}}

\usepackage{graphics}

\newcommand{\be}{\begin{eqnarray}}
\newcommand{\ee}{\end{eqnarray}}

\def\bea{\begin{eqnarray}}
\def\eea{\end{eqnarray}}
\usepackage{epsfig}

\usepackage{color}

\begin{document}
\title{$SU(5)$ Octet Scalar at the LHC}
\author{S. Khalil$^{1,2}$, S. Salem$^{1,3}$, M. Allam$^3$}
\vspace*{0.2cm}
\affiliation{$^1$ Center for Theoretical Physics, Zewail City for Science and Technology, 6 October City, Cairo, Egypt.\\
$^2$Department of Mathematics, Faculty of Science,  Ain Shams University, Cairo, Egypt.\\
$^3$Department of Physics, Faculty of Science, Al Azhar
University, Cairo, Egypt. }

\begin{abstract}
Color scalars are salient features of non-minimal $SU(5)$ model, where the Higgs sector is extended by $45$-dimensional multiplet. We show that the gauge coupling unification can be realized in this model with TeV octet scalars and intermediate scale ($\sim 10^8$ GeV) color-triplet scalars at $10^{16}$ GeV. We also analyze possible LHC signatures of these TeV octet scalars. We show that multi-(b)-jet final states provide significant signal for a direct probe of the octet scalars at the LHC.

\end{abstract}
\maketitle

\section{Introduction}

The major goal of the Large Hadron Collider (LHC) experiments at
CERN is to probe new physics beyond the Standard Model (SM) at TeV
energy scales. The grand unified theory based on $SU(5)$ gauge
symmetry is one of the most appealing scenarios for possible
extension of the SM. The minimal $SU(5)$ accommodates the matter
fields in $ 5^*$ and $ 10$ dimensional representations, while the
scalar sector consists of 24 and 5 dimensional Higgs multiplets \cite{Georgi:1974sy}.

However, minimal $SU(5)$ suffers from several severe problems.
For instance, it does not seem to unify the SM gauge couplings. In
addition, it predicts wrong fermion mass relations:
$m_{\mu(e)}=m_{s(d)}$ that contradict the experiment
measurements. A possible approach to overcome some of these
problems, is to introduce an extra Higgs multiplet with $45_H$
dimensional
representation \cite{Frampton,P5,Khalil:2013ixa}. The $45_H$ transforms under the SM gauge as%
\bea %
45_H = (8,2,1/2)\oplus (1,2,1/2)\oplus (3,1,-1/3)\oplus
(3,3,-1/3) \oplus (6^*,1,-1/3)\oplus (3^*,2,-7/6)\oplus
(3^*,1,4/3).%
\eea%
It also satisfies the following constraints: $45^{\alpha
\beta}_\gamma = - 45^{\beta \alpha}_\gamma$ and $\sum_\alpha^5
(45)^{\alpha \beta}_\alpha =0$. Thus, the electroweak symmetry
$SU(2)_L \times U(1)_Y$ can be spontaneously broken into
$U(1)_{em}$ through non-vanishing Vacuum Expectation Values (VEVs) of $5_H$
and $45_H$, namely%
\bea%
\langle 5_H \rangle = v_5,~~~~~~ \langle 45_H \rangle^{15}_1 &=& \langle 45_H \rangle^{25}_2 =
\langle 45_H \rangle^{35}_3 = v_{45},~~~~~~  \langle 45_H
\rangle^{45}_4 = -3 v_{45}.%
\eea
The $45_H$-doublet is defined as in \cite{Khalil:2013ixa}%
\bea \nonumber D^a \equiv (1,2)_{1/2} = \delta_b^c (45)^{ab}_c = - \delta_j^i (45)^{a i}_j \equiv  \left(
                        \begin{array}{c}
                          D^+ \\
                          D^0 \\
                        \end{array}
                      \right). \ee
While, $45_H$-color octet scalars are given by \cite{P5}%
\bea%
S^{ia}_j \equiv (8,2)_{1/2} = (45_H)^{ia}_{j} -
\frac{1}{3}\delta^i_j (45_H)^{ma}_{m} = \left(
                                 \begin{array}{c}
                                   S^{+} \\
                                   S^{0}_R+iS^{0}_I \\
                                 \end{array}
                               \right) \equiv S^A T^A, %
\eea%
where $i,j= 1,2,3$, $A= 1, . . , 8$, and $T^A$ are the $SU(3)$
generators. It is clear that the octet scalars are defined as such that
they have vanishing VEVs. Moreover, one can define the other
components of the $45_H$ as follows:
\bea
(6^*,1,-1/3) &\equiv&  \phi_{lk} = \epsilon_{lij}45^{ij}_{k} ,  \\ 
(3^*,2,-7/6) &\equiv&  \phi_{lc}= \epsilon_{lij} 45^{ij}_{c},  \\
(3^*,1,4/3) &\equiv&  \phi_k = \epsilon_{ab} 45^{ab}_{k},\\
 (3,1,-1/3) &\equiv& T_1^i = \delta^c_b  45^{ib}_c = - \delta^k_j 45^{ij}_k, \\
(3,3,-1/3)  &\equiv&  T^i_3 =  45^{ib}_c -\frac{1}{2} \delta^b_c 45^{id}_d .
\eea

These scalars acquire masses from the potential $V(5_H, 24_H,
45_H)$ after breaking the $SU(5)$ into $SU(3)_C \times SU(2)_L
\times U(1)_Y$ through the VEV of
$24_H$ and the electroweak symmetry through the VEV of the doublets
of $5_H$ and $45_H$. Therefore, some of them could be light if one
considers a possible fine tuning similar to the famous
doublet-triplet splitting in $5_H$ \cite{DT_su5(2011)}.  
We will assume that the scalars $S^{ia}_j$, $D^{a}$, and $T^i_3$
are light with masses of order TeV, while the rest of the 45- scalars 
are superheavy \cite{P5}; {\it i.e.}, their masses
are of order $\langle 24_H \rangle$ that breaks $SU(5)$ to SM at
GUT scale. It is worth noting that in our model  \cite{Khalil:2013ixa} light triplet $T_3^i$ does not contribute to the proton decay since it has no coupling with
$QQ$ \cite{int}, due to the assumption of a symmetric Yukawa coupling for the interaction: $(10 ~10 ~45)$ .

The aim of this article is twofold. Firstly, to show how modifying the particle 
content in the effective SU(5) theory realize the gauge couplings unification at $10^{16}$ GeV, that meets
the theoretical predictions of the GUT energy scale \cite{unify}. Secondly, to study possible signature of the light colored scalars at the LHC. The paper is organized as follows. In section 2 we study the gauge coupling unification in this class of models. Section 3 is devoted to direct searches for (neutral and charged) octet scalars at the LHC. Finally, we give our conclusions in section 4.

\section{Gauge Coupling Unification}
In quantum field theory, gauge couplings are functions of the
energy at which they are measured and their Renormalization Group
Equations (RGE), at one loop, are given by
\begin{equation}
\frac{d \alpha_i(t)}{d t}= \frac{b_i}{2\pi} \alpha_i^2(t) ,
\hspace{1cm} i=1,2,3 \label{rge}
\end{equation}
where $t = \ln \mu $ with $\mu$ is the running scale from the Electroweak scale, $\mu_Z$, up to
the scale of grand unification theory, $\mu_{\mathrm{GUT}}$. The couplings $\alpha_i$ are defined as $\alpha_i(t)=
\frac{g_i^2(t)}{4\pi}$ and the $U(1)_Y$ coupling $\alpha_1$ is
normalized by the factor $5/3$. One can solve these RGEs to obtain
$\alpha_i(\mu)$ at a scale $ \mu$ for given $\alpha_i(\mu_0)$,
\begin{equation}
\alpha_i(\mu)^{-1} = \alpha_i(\mu_0)^{-1}-\frac{b_i}{2 \pi}
\ln(\frac{\mu}{\mu_0}). %
\label{soln}
\end{equation}
The one loop coefficients $b_i$ are defined as %
\be%
b =- \frac{11}{3} T_{G}(R) + \frac{2}{3} T_F (R) + \frac{1}{3} T_B(R), %
\label{bG}\ee%
where $G, F, B$ stand for gauge bosons, fermions and bosons
respectively. The Dynkin index  $T(R)$ is defined for a
representation $R$ as  $T(R) \delta_{ab} = Tr(T_a T_b),$ where
$T_a$ are the generators of this representation. In $SU(N)$,
$T({\rm vector})= \frac{1}{2}$ and $T({\rm adjoint})=N$.
Therefore, the $b_i$ in the SM are given by%
\be %
b_1^{\rm SM} = \frac{41}{10},~~~~ b_2^{\rm SM} = -\frac{19}{6}, ~~~~ b_3^{\rm SM} = -7.%
\ee%
In this case, one can easily show that using the experimental
values of the coupling constants at electroweak scale:
$\alpha_{em}^{-1} = 127.9, \sin^2\theta_W= 0.2329$ and
$\alpha_3 = 0.110$, the three coupling constants will never meet
at single point.

If one assumes that the mass of $SU(5)$-Higgs doublet is of order $\mu_Z \sim {\cal O}(100)$ GeV, while the octet scalar masses are
of order $\mu_S \sim {\cal O}(1)$ TeV and the triplet scalar
masses are of order $\mu_T \gg \mu_S$, then the following
three regions for gauge coupling evolution are obtained: From $\mu_Z$ to
$\mu_S$, from $\mu_S$
to $\mu_T$ and from $\mu_T$ to $\mu_{\rm GUT}$.  In these regions the one loop $b_i$- coefficients are given by%
\bea%
b_i^{\small{\rm EW}} &= &\left(\frac{21}{5}, -3, -7
 \right),~~~~
b_i^{\small S} = \left(5, -\frac{5}{3}, -5
 \right),~~~~ b_i^{\small T} = \left(\frac{26}{5}, \frac{1}{3},
 -\frac{9}{5}
 \right).%
\eea%
Therefore the RGE solution in Eq. (\ref{soln}) takes the form
\bea%
\alpha_i^{-1}(\mu_Z) = \alpha_{_{GUT}}^{-1} -
\frac{b_i^{^{EW}}}{2\pi} \ln \left(\frac{\mu_S}{\mu_Z}\right) +
\frac{b_i^{^{S}}}{2\pi} \ln \left(\frac{\mu_T}{\mu_S}\right)+ \frac{b_i^{^{T}}}{2\pi} \ln
\left(\frac{\mu_{_{GUT}}}{\mu_T}\right). \eea
One can eliminate $\alpha_{GUT}^{-1}$ from the above equations and
consider the following two equations in terms of the unknown
scales: $\mu_S$, $\mu_T$ and $\mu_{GUT}$:
\begin{widetext}
\bea%
\alpha_1^{-1}(\mu_Z) - \alpha_2^{-1}(\mu_Z) &=& \frac{1}{2\pi}
\left(b_2^{^{EW}} - b_1^{^{EW}}\right)\ln
\left(\frac{\mu_S}{\mu_Z}\right) + \frac{1}{2\pi} \left(b_2^{S} -
b_1^{S}\right)\ln \left(\frac{\mu_T}{\mu_S}\right) +
\frac{1}{2\pi} \left(b_2^{T} - b_1^{S}\right)\ln
\left(\frac{\mu_{_{GUT}}}{\mu_T}\right), \\
\alpha_2^{-1}(\mu_Z) - \alpha_3^{-1}(\mu_Z) &=& \frac{1}{2\pi}
\left(b_3^{^{EW}} - b_2^{^{EW}}\right)\ln
\left(\frac{\mu_S}{\mu_Z}\right) + \frac{1}{2\pi} \left(b_3^{S} -
b_2^{S}\right)\ln \left(\frac{\mu_T}{\mu_S}\right) +
\frac{1}{2\pi} \left(b_3^{T} - b_2^{S}\right)\ln
\left(\frac{\mu_{_{GUT}}}{\mu_T}\right). %
\eea%
\end{widetext}
\begin{figure}[!ht]
\begin{center}
\includegraphics[scale=0.3]{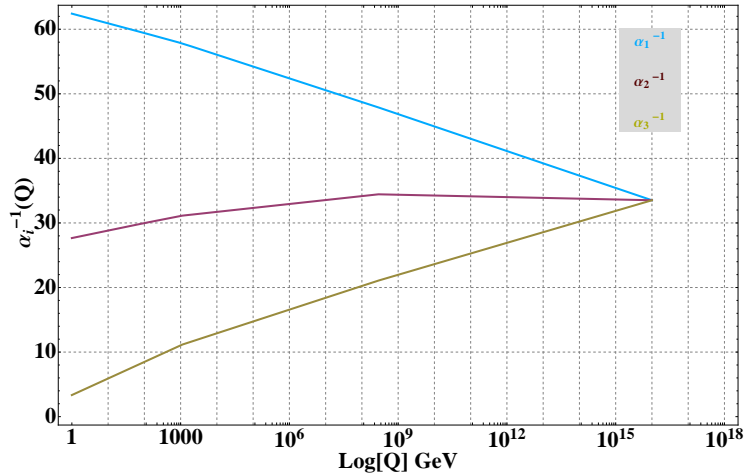}
\caption{\label{uni4} The running of the gauge couplings in low
energy effective $SU(5)$ model with electroweak scale 2HD, TeV scale octet scalars, and
intermitted scale ($\sim 10^8$ GeV) triplet scalars.}
\end{center}
\end{figure}

From these equations, and as shown in Fig. \ref{uni4}, one finds that the three gauge couplings are
unified at $\mu_{GUT} = 10^{16}$ GeV if $\mu_s$ is of order ${\cal O}(1)$ TeV and $\mu_T \sim 10^{8}$ GeV. 
It is also worth noting that the recent CMS and ATLAS experimental results,
based on searches for dijet pair signatures at $\sqrt{s} =7$ TeV,
imposed stringent constraint on the octet scalar masses: $m_S
\gsim 2$ TeV \cite{ATLAS:2012ds}.
%
\section{Octet scalar at the LHC}

TeV scale octet scalars can be produced copiously at the LHC and
lead to very interesting signatures. In the rest of the paper, we
discuss the phenomenology of these scalars, in particular their
production and decay in the LHC.
The
$SU(5)$ invariant
Yukawa Lagrangian is given by%
\bea%
{\cal L}_{\rm Yuk}= Y_1 \bar{5}_{\alpha} 10^{\alpha\beta}
(5^*_{H})_\beta + Y_2 \bar{5}_{\delta} 10^{\alpha\beta}
(45^{*}_H)_{\alpha \beta}^\delta +
\epsilon_{\alpha\beta\gamma\delta \lambda} \Big[Y_3
10^{\alpha\beta} 10^{\gamma\delta} 5_H^\lambda + Y_4
10^{\alpha\beta} 10^{\xi \gamma}_L (45_H)^{\delta \lambda}_\xi
\Big].~%
\label{yuk}
\eea%
After $SU(5)$ symmetry breaking, the interaction lagrangian of the colored octet scalars with SM
fermions can be derived as \cite{Khalil:2013ixa}
\be %
{\cal L}_{int} = 2 (Y_2)_{ij} \bar{d}_{Ri} Q_{Lj} S^{\dagger} + 4
\epsilon_{\alpha \beta} ( Y^T_4-Y_4)_{ij} \bar{u}_{Ri} Q^\alpha_{Lj}
S^{\beta}+ h.c. , ~~ %
\label{LS} %
\ee %
where the Yukawa couplings $Y_2$ and $Y_4$, along with the Yukawa couplings $Y_1$ and $Y_3$ of $5$-plet Higgs, define the fermion masses as follows  \cite{Georgi:1979df,Nandi:1980sd} %
\bea%
M_E &=& Y_1^T  v^*_5 - 6 Y_2^T v^*_{45},\label{mE}\\
M_D &=& Y_1 v^*_5 + 2 Y_2 v^*_{45}, %
\label{mD}
\\
M_U &=& 4 (Y_3 + Y_3^T) v_5 - 8 (Y_4^T - Y_4) v_{45}.%
\label{mU}%
\eea %
Therefore, the Yukawa coupling $Y_2$ can be written in terms of charged
lepton and down quark masses
$$Y_2=\frac{M_D-M_E}{8v_{45}}.$$
However, the situation for $Y_4$ is more involved. It is clear that one cannot relate $Y_4$ directly to the up-quark masses. Moreover, in the basis where $M_U$ is diagonal, {\it i.e.}, the quark mixing is  emerging from down quark sector only and the rotational matrices are given by $V_L^u =V_R^u=I$ and $V_L^d= V_{CKM}$, then $Y_3$ and $Y_4$ matrices should be also diagonal, unless there is a significant fine tuning between them. In this case one finds that  the couplings of up-quarks with octet scalars vanish identically. This conclusion can be also obtained if $Y_4$ is a symmetric matrix. Thus, the interaction Lagrangian in the physical mass basis is given by%
\begin{widetext}
\bea%
{\cal L}^S &=& \bar{d} \Big[ P_L \frac{(m_D V^\dag_{_{CKM}}-
m_E)}{4\upsilon_{45}}\Big]S^- u
 + \bar{u}\Big[ P_R \frac{( V_{_{CKM}}\ m_D-m_E)}{4\upsilon_{45}} \Big] S^+ d
- \frac{S_R^0}{4\upsilon_{45}}\Big[\bar{d} \Big(P_L (m_E V_{_{CKM}}) + P_R (V^\dag_{_{CKM}} m_E)\Big) d \nonumber \\
&-&  m_D \bar{d}{d} \Big] - \frac{iS_I^0}{4\upsilon_{45}}\Big[\bar{d} \Big(P_L (m_E V_{_{CKM}}) - P_R (V^\dag_{_{CKM}} m_E)\Big) d -  m_D \bar{d}\gamma_5{d} \Big]. \label{LS2}
 \eea
\end{widetext}
Furthermore, the gluon interaction with the octet scalars is one of their most relevant interactions with the SM particles.
It is given by \cite{Khalil:2013ixa}%
\bea %
\nonumber  {\cal L}^S_{gluon} &=& i g_s {\rm Tr}\Big[ S^{A-}
G^{\mu B} \partial_\mu  S^{D+} + S^{A0}_R G^{\mu B}
\partial_\mu  S^{D0}_R + S^{A0}_I\ G^{\mu B}  \partial_\mu  S^{D0}_I \Big]{\cal F}^{ABD} \nonumber\\
&+& g_s^2 {\rm Tr}\Big[ S^{A-}  G^{\mu B}  G^{C}_\mu S^{D+} +
S_R^{A0} G^{\mu B}  G^{C}_\mu  S^{D0}_R  + S^{A0}_I  G^{\mu B}
G^{C}_\mu  S^{D0}_I \Big]{\cal F}^{ABCD}+ h.c , ~~~%
\eea%
where %
\bea%
{\cal F}^{ABD}&=& tr[T^A T^B T^D] = 1/4 \Big( d^{ABD} + i f^{ABD}
\Big) ,\\%
{\cal F}^{ABDE} &=& tr[T^A T^B T^D t^E] =\frac{2}{9} \delta^{AB}
\delta^{DE} +\frac{1}{8} \Big[ d^{ABC} d^{DEC} +i d^{ABC} f^{DEC}
+i f^{ABC} d^{DEC} - f^{ABC} f^{DEC} \Big],%
\eea %
with $d^{ABC}$ and $f^{ABC}$ are the $SU(3)$ symmetric and
antisymmetric structure constants, respectively.

\begin{figure}[t]
\begin{center}
\includegraphics[scale=0.58]{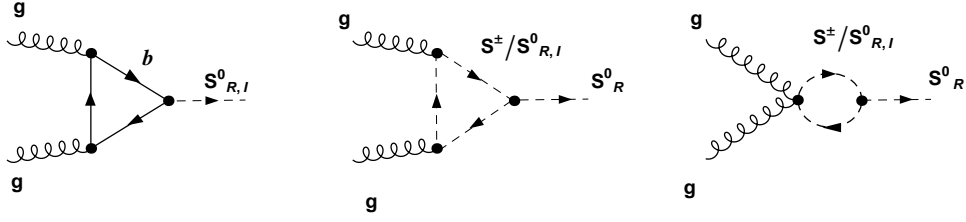}
\vspace*{-12cm}
\caption{Diagrams of the neutral octet scalar single production via gluon fusion.}
\label{fig21}
\end{center}
\end{figure}

We now consider the searches for the octet scalars at the LHC. The single neutral octet scalars can be produced at tree level from quark-anti-quark annihilation: $q \bar{q} \to S^0_{R,I}$, with $q=d,s$, and also at one loop level through the gluon fusion process: $gg \to S_{R,I}^0$, with $b$-quark or $S^{\pm}$/$S^0_{R,I}$ exchanges as shown in Fig.\ref{fig21}. for $m_{S} \gsim 2$ TeV, one can safely neglect the octet scalars contributions in the loops. Thus, the gluon fusion cross section for the neutral octet scalars is given by%
\bea
&& \hat{\sigma}_{LO}(gg \rightarrow S^0_{R,I}) = \sigma^S_0\  m^2_{S}\  \delta(\hat{s}- m^2_{S}), \\&&
\sigma^S_0 =\frac{\alpha_s^2 }{256 \pi} \Big|{\cal F}^{ABC} \frac{Y_{Sb\bar{b}}}{m_b}   A_{1/2}(\tau_b)\Big|^2, \eea%
where%
\be A_{1/2}(\tau) = 2 \Big[\tau +(\tau-1) f(\tau)\Big] \tau^{-2}, \ee%
\be f(\tau) =       \left\{ \begin{array}{rcl}
 \arcsin^2 \sqrt{\tau} ~~~~~~~~~~~~~~~~~~~~~~~~~~
 \tau \leq 1 \\  -\frac{1}{4}\Big[ \log \frac{1+\sqrt{1-\tau^{-1}}}{1-\sqrt{1-\tau^{-1}}} -i \pi \Big]^2   ~~~~~~~~~~\tau > 1.
       \end{array}\right.
 \ee%
The color factor;%
\be \sum_{ABC} {\cal F}^{ABC} {\cal F}^{ABC *} =\frac{112}{3}.
\ee%
The proton- proton cross section at LO is given by:
\be
\sigma (pp \rightarrow S^0_{R,I}) = x_0~ \int^1_{x_0}\frac{dx}{x}\ f(x) f(m_{S}^2/sx)~ \sigma^{S}_0(gg \rightarrow S^0_{R,I}), \nonumber
\ee%

\begin{figure}[t]
\begin{center}
\includegraphics[scale=0.58]{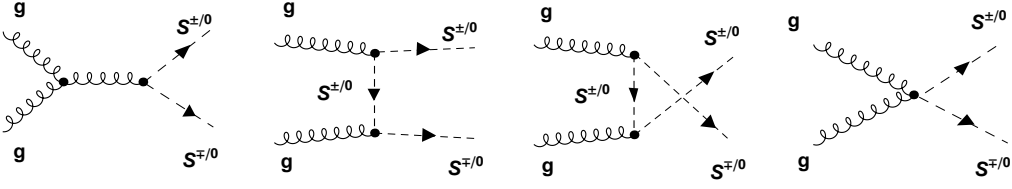}
\vspace*{-10.5cm}
\caption{Diagrams of the neutral and charged octet scalar pair production.}
\label{fig2}%
\end{center}%
\end{figure}
\begin{figure}[t]
\begin{center}
\includegraphics[scale=0.4]{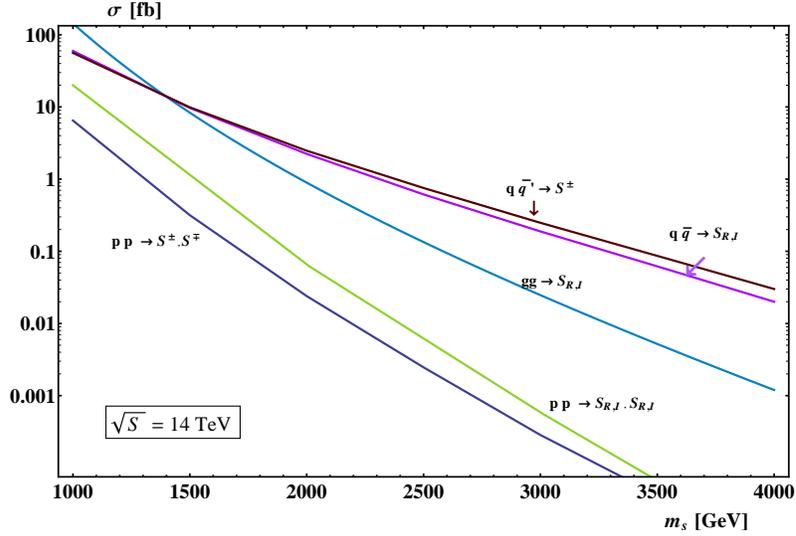}
\caption{The hadronic production cross sections of $q \bar{q} \to S^0_{R,I}$, $q \bar{q'} \to S^\pm$,  gluonic fusion
 $(gg \rightarrow S^0_{R,I})$, and the pair production $pp \to S^{0/\pm} S^{0/\mp}$ as a function of $m_{S}$ at CME $\sqrt{s}= 14~TeV.$}
\label{fig3}
\end{center}
\end{figure}

%

with $\tau= \frac{m_S^2}{4m_b^2}$, and $x_0= m_{S}^2/s$. The charged octet scalars are singly produced at tree level only from $q \bar{q'} \to S^\pm$, with $q=u,d$ or $s$.
While the pair productions of the neutral and charged octet scalars occur at tree level is shown in Fig.\ref{fig2}.
The cross section of the partonic pair production of octet scalars is given by \cite{sigma_pair}%
\bea%
\frac{d\sigma}{dt}(g g \to SS) =
\frac{\pi\alpha_s^2}{s^2} \big(\frac{27}{32} + \frac{9(u-t)^2}{32s^2} \big)
\big(1 + \frac{2m_s^2}{u -m_s^2} + \frac{2m_s^2}{t -m_s^2} \big) + \frac{2m_s^4}{(u -m_s^2)^2} + \frac{2m_s^4}{(t -m_s^2)^2}+ \frac{4m_s^4}{(t -m_s^2)(u -m_s^2) },
\eea%
which implies that
\be%
\sigma(gg \to ~SS) &=& \frac{\pi\alpha_s^2}{s}  \Big(\frac{15k}{16}+ \frac{51km_s^2}{8s} +
\frac{9m_s^2}{2s^2} (s-m_s^2)~\ln \big(\frac{1-k}{1+k}\big) \Big),
\ee
where $k=(1-\frac{4m_s^2}{s})^{1/2}$. Note that the initial kinematics threshold of this process is given by $s = 4m_S^2$.
We have used Feynrules \cite{FeynRules} to generate the model files and MadEvent5 \cite{MadG} to calculate the numerical values of the cross sections of neutral and charged octet scalar. We assume $\sqrt{s}= 14~TeV$. In Fig. \ref{fig3} we show different cross sections of single and pair productions of neutral and charged octet scalars in terms of universal octet scalar mass.
As can be seen from this figure that for octet scalar mass of order ${\cal O}(2)$ TeV, the single production cross section is about one order of magnitude larger than the pair production cross sections. This result is consistent with the findings in other extensions of the SM with heavy octet scalars \cite{P3,P4}. Therefore, one may expect that the single production of octet scalar at the LHC would have a higher possibility.

In our model, the neutral scalars decay dominantly into $b\bar{b}$, while the charged octet scalars decay into $b\bar{t}$ or $t\bar{b}$. If $m_{S^0} > m_{S^\pm}$, then one may consider the decay channel: $S^0 \to S^{\pm}~ W^{\mp}$.  Recall that the octet scalar interactions with electroweak gauge bosons are obtained from the kinetic term of $45_H$\cite{Khalil:2013ixa}: ${\rm Tr}\big[(D_\mu 45_H)^\dagger (D^\mu 45_H)\big]$
\bea
\nonumber   {\cal L} &=&
  -\frac{i m_z}{v} S^- A^\mu  \partial_\mu S^+ + \frac{ig}{\sqrt{2}} S^- W^{\mu+} \partial_\mu S^0 - \frac{ig}{\sqrt{2}} S^0 W^{\mu-}
  \partial_\mu S^+ + \frac{i m_z}{v} S^0 Z^\mu \partial_\mu S^0 \nonumber\\
  &-&  1/4\Big[\frac{ m_z^2}{v^2}  S^- A^\mu A_\mu S^+ -2 \sqrt{2} \frac{g m_z}{\nu}  S^- A^\mu  W_\mu^+ S^0 - 2g^2 S^- W^{\mu +} W_\mu^- S^+
  + 2 \sqrt{2} \frac{g m_z}{v} S^- W^{\mu +} Z_\mu S^0 \nonumber\\
  &-&  2 \sqrt{2} \frac{g m_z}{v} S^0 W^{\mu-} Z_\mu S^+ - 2g^2 S^0 W^{\mu-} W_\mu^+ S^0 + 2 \sqrt{2} \frac{g m_z}{v} S^0 Z^\mu W_\mu^-  S^+  - 4 \frac{m_z^2}{v^2} S^0 Z^\mu Z_\mu S^0\Big].
  \label{LW}  \eea

It is important to mention that phenomenological analyses of octet scalars has been recently studied in literature \cite{P4}. However, most of these analysis were based on the octet scalars given in Manohar-Wise model \cite{Manohar_model}, where the octet scalars have free coupling with SM up and down quarks as well as the SM gauge bosons. Here our octet scalars interactions are limited to the effective non-minimal $SU(5)$ model as explained in detail in Ref.\cite{Khalil:2013ixa}. Moreover, we also impose the stringent experimental constraint that neutral/charged octet scalars mass $ \sim 2$ TeV, which is consistent with our conclusions of the previous section for having a gauge coupling unification at scale larger than $10^{15}$ GeV.

\begin{figure}[!ht]%
\begin{center}
\vspace*{-0.3cm}
\includegraphics[scale=0.4]{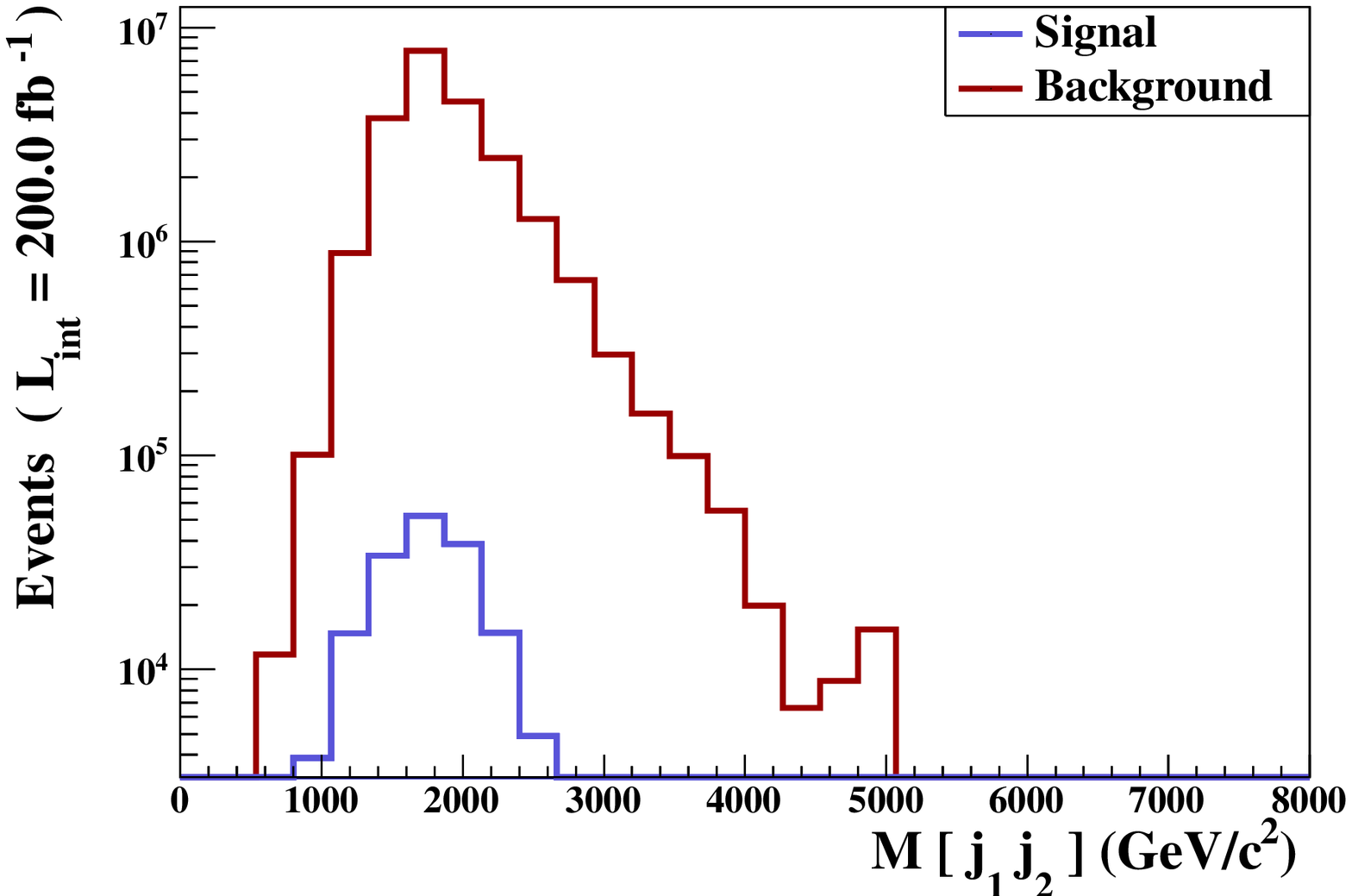}
\includegraphics[scale=0.4]{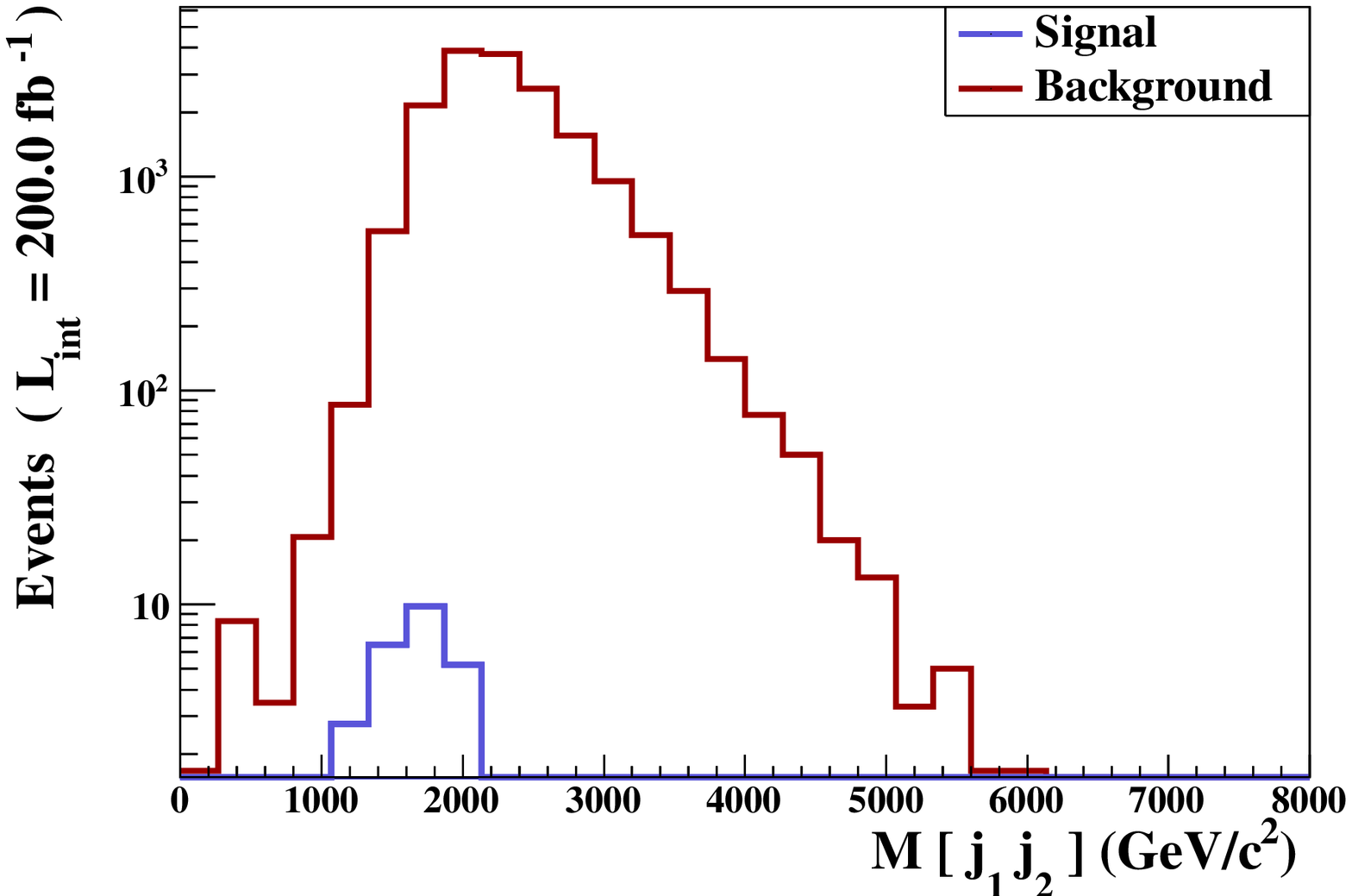}
\caption{\label{2b&ws}
The number of reconstructed events per bin of the invariant mass of a jet- pair for the $S^0$ signal and the SM background
at $P_T > 800$ GeV, $m_{S^0}=2$ TeV and $\sqrt{s}=14$ TeV for (Left panel) $p p  \to S^0 \to b \bar{b}$ and for (Right panel) $pp\to S^0 \to W^- S^+ \to l^- l^+ b\bar{b}\nu\bar{\nu}$ . Here the bin size is 30 GeV. }
\end{center}
\end{figure}

As advocated above, the process $pp \to S^0 \to b \bar{b}$ has a large cross section, however, it turns out that the SM background of this process exceeds any possible signature even if one applies a large $P_T$ cuts on the outgoing jets. In Fig. \ref{2b&ws} (left panel) we plot the number of reconstructed events per bin of the invariant mass of $b\bar{b}$ of this process for $S$ signal and SM background at $P_T~ {\rm cut} > 800$ GeV with $m_{S^0}=2$ TeV and $\sqrt{s}=14$ TeV. This figure shows that it is not possible to extract a good significance for the octet signal in this channel. In addition, we also consider the process: $pp \to S^0 \to S^+ W^- \to t ~\bar{b}~ W^- \to 2l+ 2b$ + missing energy. This process has a small cross section in SM, hence applying small cuts on the jets final states suppresses the SM background. Nevertheless, these cuts also suppress our signal. Therefore, as can be seen from Fig.\ref{2b&ws} (right panel), one cannot get a good significance for $S$ signal for this channel as well.

\begin{figure}[!ht]%
\begin{center}
\vspace*{0.2cm}
\includegraphics[scale=0.4]{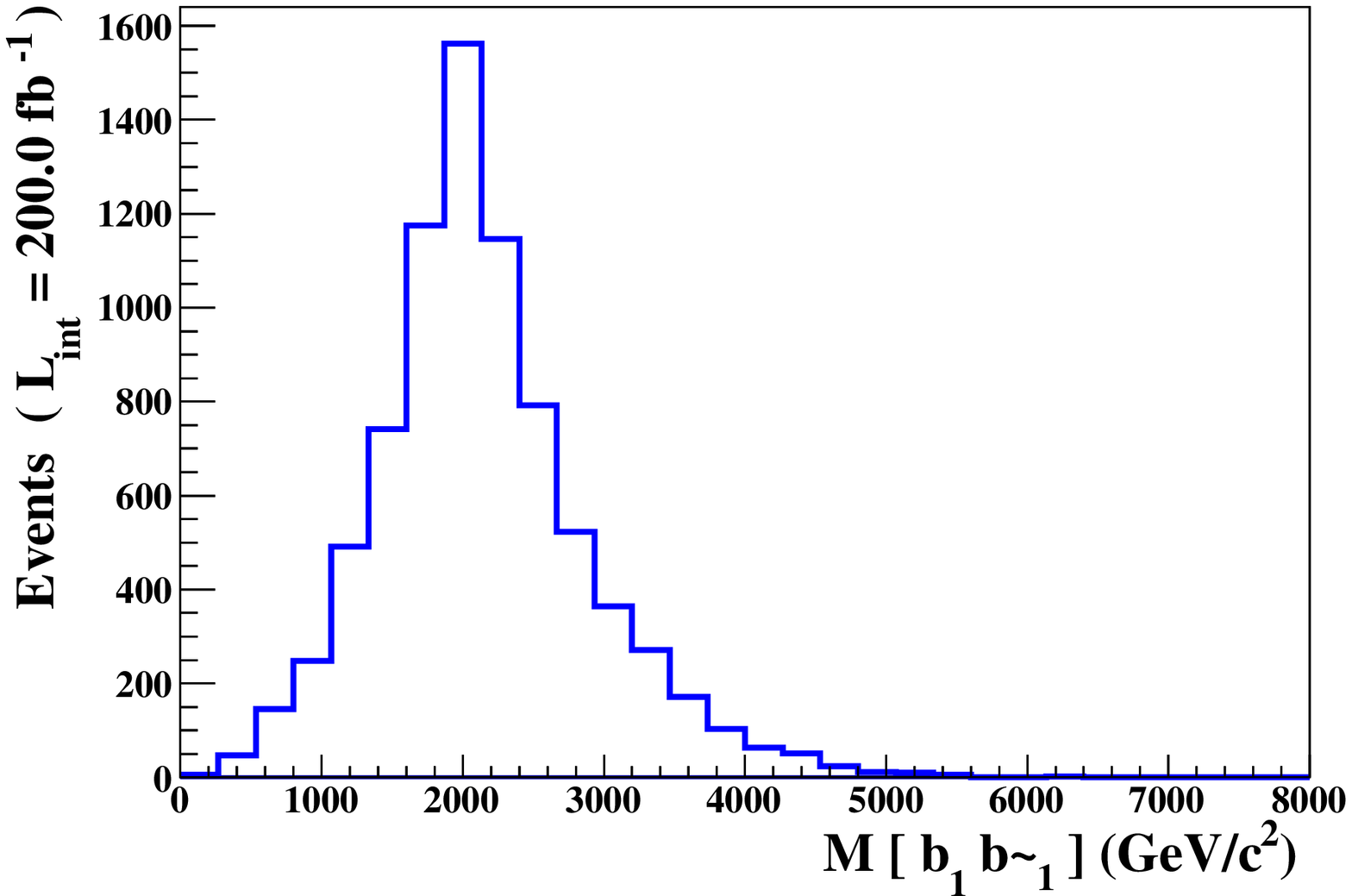}~
\includegraphics[scale=0.4]{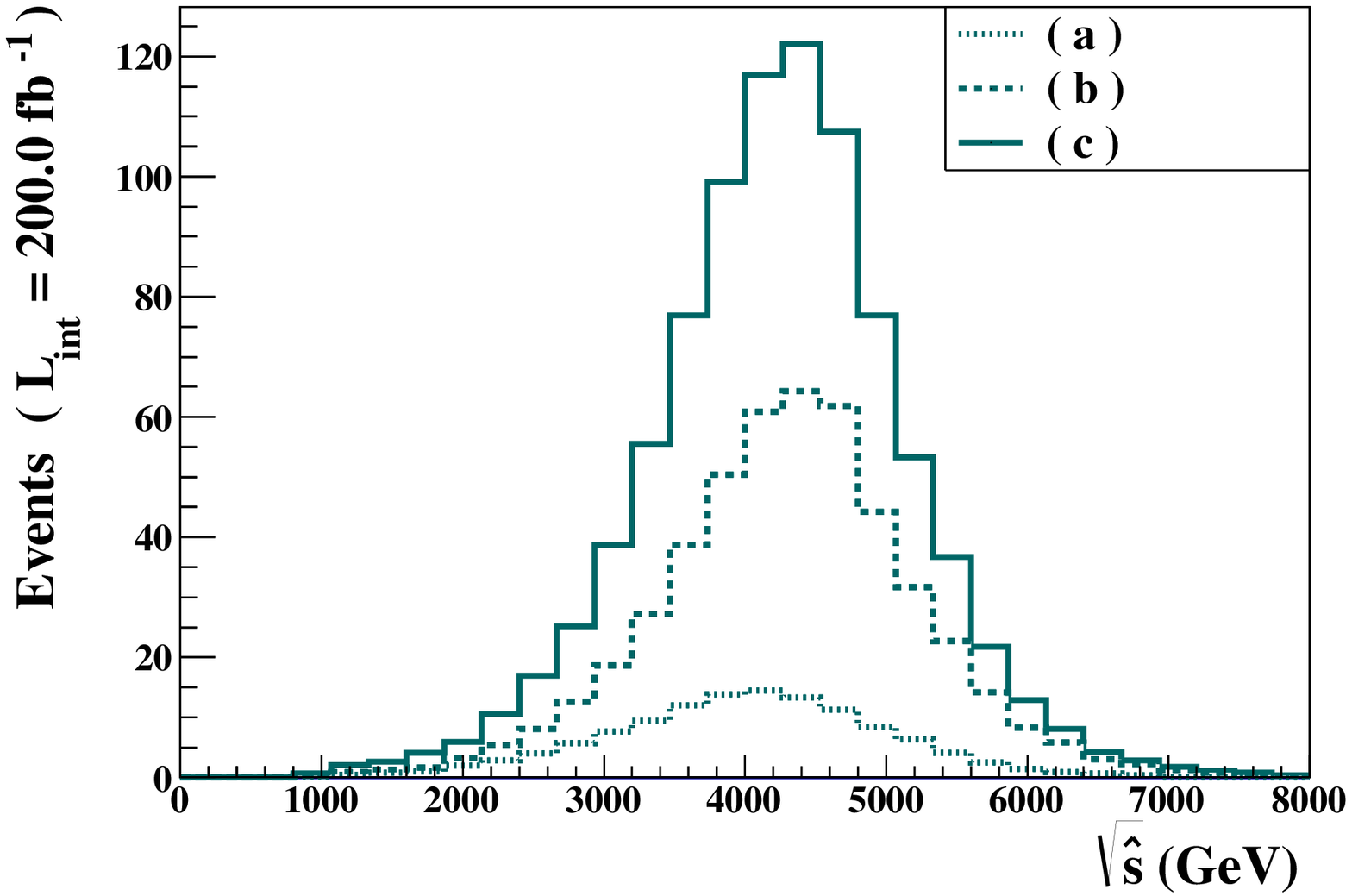}
\caption{\label{bb_Ev} (Left panel) The number of events per bin of the invarinat mass of  $b \bar{b}$ pair in $p p \to S^0 S^0 \to b~ \bar{b}~ b~ \bar{b}$\ process at $\sqrt{s} = 14 TeV$. (Right panel)The partonic center of mass energy for
$~p p \to S^+ S^- \to b~ \bar{t}~ t~ \bar{b}$ (Left panel) and  $p p \to S^0 \to S^+ ~W^-$ (Right panel) according to the decay chains~ (a),(b) and (c) for each channel as declared in the text. The size of the bin is 30 GeV and both plots
are presented without applying any cuts on the outgoing final states.}
\vspace*{-0.4cm}
\end{center}
\end{figure}

We now turn to possible signals from the pair production processes. Although, these processes have smaller cross sections, one finds that by imposing suitable cuts a significant signal can be obtained. In Fig. \ref{bb_Ev} (left panel) we plot the number of events of $p p \to S^0 S^0 \to b~ \bar{b}~ b~ \bar{b}$ per bin at the parton level at 14 TeV center of mass energy versus the invariant mass of the $b\bar{b}$ pair without applying any cuts on the outgoing b$-$jets. While in Fig. \ref{bb_Ev} (right panel), we present the partonic center of mass energy of the process $p p \to S^+ S^- \to b~ \bar{t}~ t~ \bar{b}, (\bar{t} \to W^-~ \bar{b}),(t \to W^+ b$), with the possibility that $W$ boson decays leptonacilly into $l \nu$ (30 $\%$) or hadronically into $q~\bar{q}$ ~(70 $\%$). So the following three modes are available:
$(a)$ ($W^- \to l^- \bar{\nu}),( W^+ \to q \bar{q'} )$ with final states, $l + 2(b \bar{b}) + $ 2 jets + missing energy,~
$(b)$ ($W^- \to l^- \bar{\nu}),(W^+ \to l^+ \nu) $ with final states, $2l+2 (b\bar{b})$ + missing energy and
$(c)$ ($ W^- \to q \bar{q'}),(W^+ \to q ~ \bar{q'})$. Also we didn't impose
any cuts on the final states in this plot.

\begin{figure}[!ht]
\begin{center}
\includegraphics[scale=0.5]{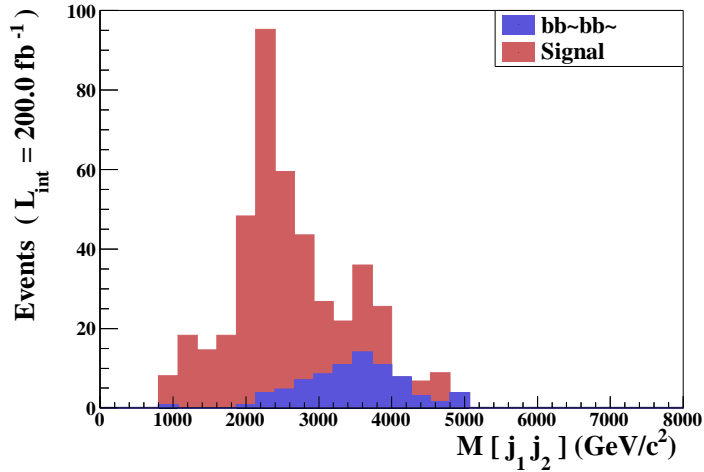}
\vspace*{-0.1cm}
\caption{\label{Sbb+back} The number of reconstructed events per bin of the invariant mass of a jet- pair for the $S^0$ signal and the SM background for $p p \to S^0 S^0 \to b~ \bar{b}~ b~ \bar{b}$ process at $P_T > 800$ GeV, $m_{S^0}=2$ TeV and $\sqrt{s}=14$ TeV. The bin size is 30 GeV  .}
\end{center}
\end{figure}

From these figures, it is clear that the $b\bar{b}b\bar{b}$ final state channel gives the best signal for probing the neutral octet scalar at the LHC.
To investigate the $4b$-tagged jets, a strong cut should be imposed to suppress the SM background.
We apply cut on the transverse momentum $P_T$ of the produced 4-jets
to be $P_T ~> 800 $ GeV. This is an acceptable cut, since we expect high energitic jets produced by the octet scalars.
This cut enhances the signifcance $S=N_{signal}/\sqrt{N_{background}}$, where $\sigma_{SM}$
becomes $3.967\times10^{-4} ~fb$ and $\sigma_S = 1.827\times10^{-3} ~fb$. Decreasing the
$P_T$ cut to $> 700 $ GeV implies a reduction in the significance of our signal, since
$\sigma_{SM}$ becomes $1.66\times10^{-3} ~fb$ and $\sigma_S = 3.954\times10^{-3} ~fb$.
In Fig. \ref{Sbb+back} we display the number of reconstructed events per bin of the invariant mass of b$-$jet pair in $p p \to S^0 S^0 \to b~ \bar{b}~ b~ \bar{b}$ process and the SM background after applying highest $P_T$ cut, $P_T >800 $ GeV at 14 TeV center of mass energy.
Here we use pythia and PGS detector simulator for the octet scalars signal and SM background at integrated luminosity $200 fb^{-1}$ and $\sqrt{s}=14$ TeV \cite{lumi_14TeV}. For jet clustering algorithm, cone algorithm with radius parameter $R =0.5$ has been used. We also used MadAnalysis5 \cite{MadA} to plot our results.

In Table \ref {ta} we provide some details for the used cut on $P_T$ and $E_T$ on signal and background for the process $pp\to S^0 \to W^- S^+ \to l^- l^+ b\bar{b}\nu\bar{\nu} $.
As can be seen from the results in this table, the signal of this process is always smaller than the background. It is worth mentioning that we have tried other cuts like pseudo-rapidity, $\eta$, and we find that similar to $P_T$ and $E_T$ the signal remains less than the background.

\begin{table}[!ht]
\centering
\begin{tabular}{|c|c|c|c|}
\hline
$P_T$min.~[GeV]   & $E_T$min.~[GeV] &  $\sigma_S~ [fb]$ & $\sigma_{SM} [fb]$   \\ [0.5ex]
\hline
\hline
$800$ & $ 0 $ & $1.37\times10^{-4}$ &  $8.3\times10^{-2}$   \\
[1ex]
$600$ & $ 300 $ & $3.9\times10^{-4}$ &  $6.8\times10^{-2}$   \\
[1ex]
$800$ & $ 400 $ & $4.4\times10^{-5}$ &  $6.9\times10^{-3}$   \\
[1ex]
$200$ & $ 400 $ & $2.1\times10^{-3}$ &  $0.88$   \\
[1ex]
\hline
\hline
\end{tabular}
\caption{Different cross sections for $pp\to S^0 \to W^- S^+ \to l^- l^+ b\bar{b}\nu\bar{\nu} $ process and the SM background for
different cuts applied on the transverse momentum $P_T$ and the missing energy $E_T$  at $m_{S^0} = 2~$TeV and CME=14 TeV.  }
\label{ta}
\end{table}

\section{Conclusions}
In this paper, we have studied some phenomenological aspects of the non-minimal $SU(5)$ model. We emphasized that the low energy Higgs sector of this class of models consists of two SM-like Higgs doublets, two neutral and one charged octet scalars, in addition to two triplets colored scalars. We have shown that the gauge coupling unification could be realized at $10^{16}$ GeV, if the octet scalar masses are of order TeV and the triplet scalar masses are of order an intermediate scale $\sim 10^8$ GeV. We have also analyzed the possible LHC signatures of these TeV octet scalars. We showed that they can be singly produced at tree level from $q\bar{q}$ annihilation (neutral octet scalar) or $q\bar{q'}$ annihilation (charged octet scalar) and in pairs through the process $gg \to S^{0(\pm)} S^{0(\mp)}$. We found that the process $pp \to S^0 \to b \bar{b}$ has the largest cross section, however the SM background is quite large and exceeds any possible signature even if one applies a large $P_T$ cut on the outgoing jets. We argued that applying a high $P_T$ cut of the order $\gsim 800$ GeV on the outgoing jets, the best channel which can provide a good signal for our heavy octet scalars is the $4b$-tagged jets final states. Although the other channels of multi-jets with associated leptons and missing energy have significant cross sections, they can not lead to good significance due to the high $p_T$ cut that one should implement in order to suppress the SM background.

\section{Acknowledgment}

We would like to thank I. Dor$\hat{s}$ner for his useful comments about the previous version.
S.S. would like thank Ahmed Ali, Celine Degrande, Olivier Mattelaer, Peter Skands, Benjamin Fuks and
Rikkert Frederix for their useful help and discussions.

\bibliographystyle{plain}

\end{document}